\RequirePackage{ifpdf}
\documentclass[11pt,a4paper]{article}
\pdfoutput = 1
\usepackage{jcappub}

\usepackage{mathrsfs}
\usepackage{dcolumn}
\usepackage{bm}
\usepackage{color}

\providecommand{\beqa}{\begin{eqnarray}}
 
 \providecommand{\rm}{\mathrm}
\providecommand{\eeqa}{\end{eqnarray}}

\usepackage{ulem}

\newcommand{\beq}{\begin{equation}}
\newcommand{\eeq}{\end{equation}}
\newcommand{\e}{\epsilon}

\newcommand{\az}{\alpha_0}

\newcommand{\bz}{\beta_0}
\newcommand{\pa}{\partial}

\title{Getting Super-Excited with Modified Dispersion Relations}
\author[a]{Amjad Ashoorioon}
\author[a]{Roberto Casadio}
\author[b,c]{Ghazal Geshnizjani}
\author[b]{ Hyung J. Kim} 

\affiliation[a]{I.N.F.N., Sezione di Bologna, IS FLAG viale~B.~Pichat~6/2, I-40127 Bologna, Italy}
\affiliation[b]{Department of Applied Mathematics, University of Waterloo\\ Waterloo, Ontario, N2L 3G1, Canada}
\affiliation[c]{Perimeter Institute for Theoretical Physics\\ 31 Caroline St. N., Waterloo, ON, N2L 2Y5, Canada}

\emailAdd{amjad.ashoorioon@bo.infn.it}
\emailAdd{roberto.casadio@bo.infn.it}
\emailAdd{ggeshniz@uwaterloo.ca}
\emailAdd{h268kim@uwaterloo.ca}

\date{\today}

\abstract{We demonstrate that in some regions of parameter space, modified dispersion relations can lead to highly populated excited states, which we dub as ``super-excited'' states.  In order to prepare such super-excited states, we invoke dispersion relations that have negative slope in an interim sub-horizon phase at high momenta. This behaviour of quantum fluctuations can lead to large corrections relative to the Bunch-Davies
power spectrum, which mimics highly excited initial conditions. We identify the Bogolyubov coefficients that can yield these power spectra.
In the course of this computation, we also point out the shortcomings of the gluing method for evaluating
the power spectrum and the Bogolyubov coefficients. As we discuss, there are other regions of parameter space, where the power spectrum does not get modified. Therefore, modified dispersion relations can also lead to so-called ``calm excited states''. 
We conclude by commenting on the possibility of obtaining these modified dispersion relations within
the Effective Field Theory of Inflation. }


\begin{document}
\maketitle
\section{Introduction}
Much has been said and written about the possible effects due to perturbation modes evolving inside
the horizon with energies larger than the scale of known physics (see
Refs.~\cite{Martin:2000xs, Easther:2001fi, Easther:2001fz, Danielsson:2002kx, Easther:2002xe, Kaloper:2002uj,Kempf:2000ac, Kempf:2001fa, Chatwin-Davies:2016byj, Ashoorioon:2004vm, Ashoorioon:2004wd, Ashoorioon:2005ep,Alberghi:2006th} for a few examples).
Depending on how one models the new physics, various assessments for the amplitude of the
corrections to standard quantum field theory results were suggested.
In fact, in most scenarios, these effects were estimated to be of order $\left({H}/{M}\right)^n$,
where $H$ and $M$ are respectively the inflationary Hubble parameter and scale
of new physics, and the exponent $n\gtrsim 1$, typically. 
\par
One approach to model physics at very high momenta, is to assume that the Lorentzian dispersion
relation, $\omega^2=k^2$, is modified by higher order corrections.
Depending on whether the dispersion relation respects the adiabaticity condition when the mode
are inside the horizon, the corrections to the spectrum can be negligible~\cite{Martin:2002kt}
or dominant~\cite{Martin:2003kp, Zhu:2016srz}.
The authors of these studies used the gluing approximation to estimate the corrections.
However, comparing the outcome of this approximation in the case of Ref.~\cite{Martin:2002kt}
with the exact analytical solution, authors of  \cite{Ashoorioon:2011eg} showed that the gluing method overestimates
the correction to the power spectrum by an extra factor of $\left({H}/{M}\right)$.
Moreover, for a dispersion relation with non-adiabatic evolution, the numerical solution for the mode equation
suggests that the effect is drastically different from what the gluing technique predicts.
In Ref.~\cite{Martin:2003kp}, the gluing method indicated that the correction to the power spectrum
is given by an oscillatory function whose amplitude depends almost linearly on the amount of time
each mode spends in the non-adiabatic region.
The numerical solution on the other hand demonstrated that the exact behaviour of the power spectrum is more involved,
depending on other parameters in the problem~\cite{Joras:2008ck}.
It is often argued that effects of order one or higher on the power spectrum (see \cite{Shiu:2002kg} as an example) should be excluded
due to the backreaction of excited states in
the phase governed by the known physics~\cite{Tanaka:2000jw}.
However, as demonstrated in Ref.~\cite{Greene:2004np}, as long as the scale of new physics is different from the Planck scale, the back reaction is not very constraining for
corrections to observables like the power spectrum.
They estimated the upper bound on the second Bogolyubov coefficient to be of order
$\sqrt{\varepsilon\,\eta}\, H\, M_{\rm pl}/M^2$, where $\varepsilon$ and $\eta$ are respectively the first
and second slow-roll parameter for a given inflationary potential and $M_{\rm pl}$ is the Planck mass.
In fact, for a given inflationary model, the back reaction of excited states with large Bogolyubov coefficients,
could be counteracted by simultaneously reducing the ratio of $H$ and $M$ \cite{Ashoorioon:2013eia}. 
Further more, for a large-field model of inflation with highly excited initial condition the
new physics scale cannot be arbitrarily larger than the Hubble parameter during inflation \cite{Ashoorioon:2013eia}.
For example for the quadratic potential, the largest value allowed turns out to be
$M\simeq 21 \,H$. In this paper, we call such highly excited initial conditions as ``super-excited'' states.
These super-excited initial conditions can generate interesting observational effects.
For instance, super-excited states with $k$ dependence, induce a running in the spectral index larger than $\mathcal{O}(\epsilon^2)$ \cite{Ashoorioon:2014nta}. In another example, it was shown that the position-dependent modulations of excited states with large occupation numbers lead to the hemispherical anomaly~\cite{Ade:2015lrj}.
In a recent work, the effect of rotational symmetry-breaking excited initial conditions was used to obtain
statistically anisotropic power and bi-spectrum.
Such an effect is not yet observed in the CMB, but may have an impact for future galaxy surveys,
leaving a trace in the shape of galaxies~\cite{Chisari:2016xki}.
\par
All of the above motivates us to further investigate the microphysical origin of such super-excited
initial states.
Recalling that dispersion relations with an intermediate phase that break WKB condition can induce large corrections
to the power spectrum, we were inspired to further study the implications of modified dispersion relations for excited initial states.
In particular, we are exploring the modified dispersion relations of the following form 
\beq \label{disp6}
\omega^2(k_{\rm ph})=\bar{\beta}\, k_{\rm ph}^6-\bar{\alpha} k_{\rm ph}^4+  k_{\rm ph}^2 
\ , 
\eeq
where  $\bar{\beta}$, $\bar{\alpha} \geq 0$.
In particular, we like to explicitly show that the region of parameter space that results in larger than one modification of the
power spectrum corresponds to an initial condition with super excited state in standard picture.
To estimate the solution of the mode equation, we use both the analytical technique called gluing as well as numerical techniques.
Gluing method has been used quite extensively in the literature~\cite{Martin:2002kt} for estimating the power spectrum. In this method method, the solution to the differential equation in different regions of the domain are glued to each other by matching the functions and their first derivatives at the boundaries.
The initial condition is usually set to be adiabatic vaccum which corresponds to positive
frequency WKB mode at infinite past. However, one has to keep in mind that the precision of this method relies strongly on how well the solutions overlap or merge at the baundaries.
In the case of $\bar{\beta}=0$ and $\bar{\alpha}<0$, we have an exact analytic solution for mode equation~\eqref{disp6},
and therefore we can extract the {\it exact\/} correspondence to Bogolyubov coefficients. This allows us to test the reliability of the gluing prediction in this case. As we will show the second Bogolyubov coefficient goes to zero much faster than what the simple matching technique suggests. The discrepancy between the two methods can be as significant as order
$\left({H}/{M}\right)^2$. 
\par
For $\bar{\beta}\not=0$ and $\bar{\alpha}> 0$, one has a sixth order polynomial dispersion relation with an intermediate
phase that dispersion curve has negative slope. In this case no exact analytic solution is known for equation of motion~\eqref{disp6}.
However, matching excited mode solutions to the implicit solution given by Mathematica will provide
us the corresponding Bogolyubov coefficients. In our approach, we make sure that the matching is performed in the regime that solutions are overlapping to a high precision. 
As we will observe, these dispersion relation even in the cases where there is only one turning point in the equation (corresponding to usual Hubble crossing) can mimic excited states with particle number density as large as 80. This verifies the result of
previous studies~\cite{Ashoorioon:2014nta,Ashoorioon:2015pia,Ashoorioon:2016lrg}.
Our derivation also shows that in some regions of parameter space, 
above dispersion relation does not lead to any corrections to power spectrum.
We will identify the corresponding  ``calm excited states"~\cite{Ashoorioon:2010xg} resulting from the
evolution of these modes.
In the last section, we will comment on how dispersion relations like the one in Eq.~\eqref{disp6}
can arise in the context of the Effective Field Theory of Inflation~\cite{Cheung:2007st}.
Finally, we will conclude the paper and propose avenues for future developments. 
\section{Precision of gluing method for estimating power spectrum} 
\label{Sgluing}
As already pointed out, our main goal in this paper is to study the effects of modified dispersion relations
on the power spectrum.
Modified dispersion relations can lead to complicated Ordinary Differential Equations (ODEs)
in Foureir space that are hard to solve analytically.
A popular analytical method in the literature to estimate the solution of an ODEs,
is to use matching techniques. In this approach, one first finds the general solutions to the ODE in different domains.
Imposing an initial or boundary condition then fixes the coefficients in one region, and the coefficients
in the other regions are determined by matching the solutions at the boundaries of these regions.
This is in fact quite a powerful technique to estimate outgoing amplitudes in different areas of physics,
such as optics and quantum mechanics.
As is customary in the cosmology community, we will refer to this technique as {\it gluing method\/}.
The precision of this technique depends on how smoothly the solutions of different intervals merge to each other.
Depending on the intended accuracy, one can always divide the domain in more regions.
Nonetheless, doing that might not necessarily provide additional intuitions or advantages
compared to numerical methods.
\par
To highlight this point, we review the gluing method for the mode equation of the standard
Lorentzian dispersion relation.
This example will help us understand one source of offsets when we deal with more complicated
modified dispersion relations.
In this vanilla model, the mode equation for a massless spectator field $\phi$ on the de~Sitter
background (with scale factor $a=-1/H\,\tau$) and minimally coupled to gravity is 
\beq \label{modeeq}
u_k''+\left( k^2 -\frac{2}{\tau^2} \right) u_k
=0
\ .
\eeq
Here $u$ represents the canonical variable defined as $u \equiv a\,\phi$.
This equation is the modified Bessel equation and we can find exact solutions and the corresponding
power spectrum. Assuming Bunch-Davis initial conditions, the exact modes are given by
\beqa \label{PSdef}
u_k
=
\frac{1}{\sqrt{2}}\, e^{-i\,k\,\tau}\left(1-\frac{i}{k\,\tau} \right).
\eeqa
Substituting this solution into the power spectrum,
\beq\label{PS}
P_k^\phi
\equiv
\frac{k^3}{2\,\pi^2}
\langle \,\phi_k^2\, \rangle
=
\frac{k^3}{2\,\pi^2\,a^2} \langle \,u_k^2\, \rangle
\ , 
\eeq
one obtains that at late times ($\tau\to 0$),
\beqa \label{BDPS}
P_{\rm BD}
\equiv
P_k^{\phi~({\rm exact})}
=
\frac{H^2}{4\pi^2 }
\ .
\eeqa
In the following, we will use this value as our reference point and evaluate corrections with respect to this value.
\par
Suppose we now try to estimate the power spectrum via a gluing approach.
We divide the time domain $\tau\in(-\infty,0)$ into two regions, namely region~${\rm I}=\left(-\infty, -{\sqrt{2}}/{ k}\right]$
and region~${\rm II}=\left[-\sqrt{2}/{k},0\right)$. We then impose continuity of the solutions and their first derivatives at the
boundary point 
\beq
\tau_b\equiv -\frac{\sqrt{2}}{ k}
\ .
\eeq
In the limit $\tau\ll \tau_b$, Eq.~\eqref{modeeq} simplifies to a simple harmonic oscillator.
The particular solution of this equation that approaches the Bunch-Davies vacuum is given by
\beq
u_{k\,\rm I}
=
\frac{1}{\sqrt{2\, k} }\exp(-i\, k\, \tau)
\ .
\label{ukI}
\eeq
On the other hand, one can neglect $k^2$ in Eq.~\eqref{modeeq} for $\tau\gg \tau_b$.
Therefore, the general solution in the second region is given by
\beq
u_{k\, \rm II}
=
C_1 \tau^2+\frac{C_2}{\tau}
\ .
\label{ukII}
\eeq
The coefficients $C_1$ and $C_2$ can be determined by assuming $u_{k\,\rm I}$ and $u_{k\,\rm II}$
remain good approximations around the boundary point $\tau_b$, and requiring the continuity of the
function and its derivative at this point,
\beqa
u_{k\,\rm I}(\tau_b)
&=&
u_{k\,\rm II}(\tau_b)
\, \label{gluing-prscription1}
\eeqa
\beqa \label{gluing-prscription2}
\left.\frac{d u_{k\,\rm I}(x)}{d \tau}\right|_{\tau_b}
&=&
\left.\frac{d u_{k\,\rm II}(x)}{d \tau}\right|_{\tau_b}
\ . 
\eeqa
These conditions yield 
\beqa
C_1
&=&
\frac{1}{12} \left(\sqrt{2}+2\,i\right) e^{i\,\sqrt{2}}\, k^{3/2}
\\
C_2
&=&
-\frac{2-i\,\sqrt{2}}{3\,  k^{3/2}}\,e^{i\sqrt{2}}
\ ,
\eeqa
and the power spectrum~\eqref{PS} becomes  
\beqa
P_k^\phi
\approx
\frac{k^3 H^2}{2\,\pi^2}\, |C_2|^2
\approx
\frac{12}{9} \frac{H^2}{4\pi^2 }
\approx
1.33\, P^{\phi~({\rm exact})}_{k}
\ .
\eeqa
The discrepancy between the gluing result and the one obtained from the exact
normalized modes is almost $33\%$, which shows the inadequacy of the gluing technique in properly
estimating the amplitude of the power spectrum. 
\par
Of course, the reason for this discrepancy is that our approximate solutions in region~I ~\eqref{ukI} and
in region~II ~\eqref{ukII} are not a good description of the actual solution in the neighbourhood of the
turning point $\tau_b$.
If the regions of the validity of these solution had overlapped, then one would expect that the corresponding
power spectrum be a good approximation for the exact expression. 
In fact, similarly to tunnelling problems in quantum mechanics, one could improve the accuracy
by employing a WKB approximation in the first region and Airy functions around the turning point
$\tau_b$.
We should remark that, in inflationary backgrounds unlike most tunnelling problems, the WKB approximation is not
restored after crossing the turning point. 
To summarize, matching techniques are helpful to estimate the power spectrum, but when it comes
to small corrections, one should be aware of the limitations that come with this approximation. 
\section{Corley-Jacobson dispersion relation}
\label{Scj}
We now turn our attention to the simplest correction to Lorentzian dispersion relation, namely
the Corley-Jacobson (CJ) dispersion relation~\cite{Corley:1996ar} \footnote{Please also see \cite{Kahn:2015mla} for such modified dispersion for gravitino in the context of Effective Field Theory.}. This dispersion relation assumes an additional quartic term to the physical momenta in the Ultra Violet (UV)
regime, 
\beq 
\omega^2_{\rm ph}
=
k_{\rm ph}^2+\beta_0\, k_{\rm ph}^4
\ .
\eeq
\par 
Changing to comoving momenta, $k=a\,k_{\rm ph}$ and $\omega^2(k),
=
a^2\, \omega^2_{\rm ph}(k/a)$ 
 the new mode equation becomes
\beq
u_k''+\left (\e\, \tau^2\, k^4+k^2-{2\over \tau^2}\right )
u_k
=
0
\ ,
\label{eqCJ}
\eeq
where we also defined the parameter $\e\equiv \beta_0\, H^2$. 
This mode equation was extensively analyzed in Ref.~\cite{Ashoorioon:2011eg}.  
It was shown there that it has exact analytical solutions and the power spectrum can 
again be obtained without any approximations. 
Imposing adiabatic vacuum initial conditions, the exact modes are given by 
\begin{equation}
\label{exact-adiabatic}
u_{k}^{\rm (exact)}(\tau)
=
\frac{\exp(-\frac{\pi}{8\, \e})}{\sqrt{-2\,\e\,\tau}\, k}\,
{\rm WW}\!\left(\frac{i}{4\,\e},\frac{3}{4},-i\,\e\, k^2\,\tau^2\right)
\ ,
\end{equation}
with $\mathrm{WW}$ representing the $\mathrm{Whittaker\ W}$ function.
Computing the power spectrum~\eqref{PSdef}, one obtains
\beqa
P_k^{\phi~({\rm exact})}
=
\gamma^{\rm (exact)}\,P_{\rm BD}
\ ,
\eeqa
where the modulation of the Bunch-Davis power spectrum~\eqref{BDPS} is given by
\beq
\label{exact-mode}
\gamma^{\rm (exact)}
=
\frac{\pi\,  e^{-\frac{\pi }{4\, \e}}}{4\, \e^{3/2}\, \Gamma\!\left(\frac{5}{4}-\frac{i}{4\, \e}\right)
\Gamma\!\left(\frac{5}{4}+\frac{i}{4\,\e}\right)}
\ .
\eeq
In order to determine the Bogolyubov coefficients, we note that Eq.~\eqref{eqCJ}
reduces to the usual vanilla one in an expanding background as $x\rightarrow 0$.
Therefore, the exact solution~\eqref{exact-adiabatic} in this limit can be matched smoothly to an excited
mode,
\beq
\label{excitedmode}
u_{k\, \rm IV}
=
\frac{\sqrt{-x\,\pi}}{2}\left[\xi\, H_{3/2}^{(1)}(-x)+\rho\, H_{3/2}^{(2)}(-x)\right]
\ ,
\eeq
We can then obtain the corresponding Bogolyubov coefficients $\xi$ and $\rho$. Since the mode function~\eqref{excitedmode} itself is singular at $x= 0$,
instead we work with function $f(x)=x\,u(x)$ for which 
\beqa
f(0)
&=&
-i\, \frac{\xi-\rho}{\sqrt{2}}
\\
f'(0)
&=&
0
\\
f''(0)
&=&
-i\, \frac{\xi-\rho}{\sqrt{2}}
=
f(0)
\\
f'''(0)
&=&
-\sqrt{2}\,(\xi+\rho)
\ .
\eeqa
The Bogolyubov coefficients are then determined by evaluating the function $f$ and its third derivative at $x=0$.
We obtain
\beqa
\rho(\e)
&=&
\frac{\sqrt{\pi }\, e^{-\frac{\pi }{8\, \epsilon}}}{4\, \epsilon ^{3/2}}
\left[
\frac{8\, (-1)^{3/8}\, \epsilon^{9/4}}{\Gamma\!\left(-\frac{\epsilon +i}{4\, \epsilon }\right)}
-\frac{(-i\, \epsilon )^{3/4}}{\Gamma\!\left(\frac{5}{4}-\frac{i}{4 \,\epsilon }\right)}
\right]
\\
\xi(\e)
&=&
\frac{\sqrt{\pi } \,e^{-\frac{\pi }{8 \,\epsilon }}}{4\, \epsilon ^{3/2}}
\left[
\frac{8 (-1)^{3/8} \epsilon ^{9/4}}{\Gamma\!\left(-\frac{\epsilon +i}{4\, \epsilon }\right)}
+\frac{(-i\, \epsilon )^{3/4}}{\Gamma \left(\frac{5}{4}-\frac{i}{4\,\epsilon }\right)}\right] 
\ .
\eeqa
The number density of particles in an excited state is then given by
\beq
N_k(\epsilon)^{\rm exact}
=
\frac{(-1)^{5/8}\,e^{-\frac{\pi }{4 \epsilon }}}{32\, \pi\, \epsilon ^3}\,
\cosh\!\left(\frac{\pi }{2\, \epsilon }\right)
\left[
(-1)^{3/8}\,(i\,\epsilon )^{3/4}\, \Gamma\!\left(\frac{i-\epsilon}{4\, \epsilon }\right)
-8\, \epsilon^{9/4}\, \Gamma\!\left(\frac{5}{4}+\frac{i}{4 \epsilon }\right)
\right]
\ ,
\eeq
which goes like $25\, \e^4/64$ for small $\e$.
\begin{figure}[t]
\label{gammas}
\begin{center}
\includegraphics[angle=0, scale=1.3]{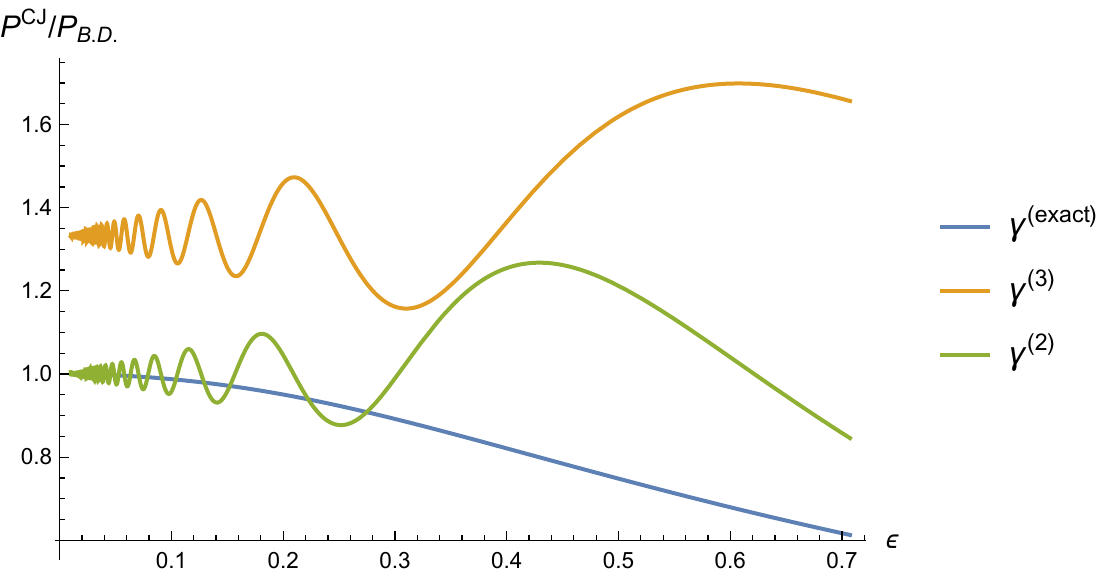}
\caption{Ratio of Corley-Jacobson power spectrum to Bunch-Davies power spectrum
from various methods: 
$\gamma^{(3)}$ is the modulation factor when using three regions for gluing solutions;
$\gamma^{(2)}$ is the modulation factor when using two regions for matching solutions;
and $\gamma^{\rm (exact)}$ is the exact modulation factor.
Notice that, beside the superimposed oscillations, further gluing at the horizon crossing
causes an offset of 1/3 with respect to the exact result around $\e=0$.
Oscillations near $\e=0$ are an artefact introduced by the gluing technique and
are absent in the exact result.}
\label{gammas-compared}
\end{center}
\end{figure}
\par
Now that we have the exact answer, we can again check the precision of the gluing technique. 
As noted previously in Ref.~\cite{Ashoorioon:2011eg}, assuming $\epsilon \ll 1/{\sqrt{2}}$,
one can decompose the time domain into three regions: 
\beqa
k\,\tau
&\leq&
-\epsilon^{-1}
\qquad 
\text{region~I}
\\
-\epsilon^{-1}
\leq
k\,\tau
&\leq&
-\sqrt{2}
\qquad
\text{region~II}
\\
k\,\tau
&\geq&
-\sqrt{2}
\qquad
\text{region~III}
\ .
\eeqa
The solution in region~I that satisfies the wronskian condition, 
\begin{equation}
\label{Wronskian}
u(\tau)\, u'^{\ast}(\tau)-u^{\ast}(\tau)\, u'(\tau)
=
i
\ ,
\end{equation}
and the adiabatic vacuum initial conditions, turns out to be 
\beq
u_{k\,\rm I}
=
\frac{
D_{-\frac{1}{2}}\!\left( (-1)^{3/4}\, \sqrt{2\, \e}\, x\right)}{{(2\,\e)}^{1/4}\,\sqrt{k}}
\ .
\eeq
Here, $D_\nu(x)$ is the parabolic cylinder function of order $\nu$, and we defined
the dimensionless parameter $x\equiv k\,\tau$.
Next, this solution is glued to the general solutions in region~II,
\beq
u_{k\,\rm II}
=
A_{2}\, \exp(i\, x)+B_{2}\,\exp(-i\, x)
\ ,
\eeq
and the coefficients are determined by requiring continuity at $x_1=-{1}/{\e}$. This yields
\beqa
A_{2}
&=&
-\frac{\left(1+i\right) e^{{i}/{\e}}
\left[\sqrt{\e}\, D_{\frac{1}{2}}\!\left(\frac{1-i}{\sqrt{\e}}\right)
-(1-i)\,D_{-\frac{1}{2}}\!\left(\frac{1-i}{\sqrt{\e}}\right)\right]}{{(32\,\e)}^{1/4}\,\sqrt{k}}
\\
B_{2}
&=&
\frac{\left(1+i\right)
e^{-i/\e}\,\sqrt[4]{\e}\,D_{\frac{1}{2}}\!\left(\frac{1-i}{\sqrt{\e}}\right)}{2\,\sqrt[4]{2}\,\sqrt{k}}
\ .
\eeqa
Likewise, matching these solutions to the general solutions in region~III,
\beq
u_{k\,\rm III}
= \frac{A_{3}}{x}+B_{3}\,x^2
\ ,
\eeq
leads to 
\beqa
A_3
=
\frac{e^{-\frac{i\left(\sqrt{2}\, \e+1\right)}{\e}}}{3\,\sqrt[4]{8\, \e}\, \sqrt{k}}
&&
\!\!\!
\left\{(1+i)\, \sqrt{\e} \left[
\left(2+i\, \sqrt{2}\right) e^{2\, i/\e} 
+i\, \left(\sqrt{2}+2\, i\right) e^{2\, i \sqrt{2}}\right]
D_{\frac{1}{2}}\!\left(\frac{1-i}{\sqrt{\e}}\right)
\right.
\nonumber
\\
&&
\left.
-\left(4+2\, i\, \sqrt{2}\right) e^{2\, i/\e}\,D_{-\frac{1}{2}}\!\left(\frac{1-i}{\sqrt{\e}}\right)
\right\}
\eeqa
\beqa
B_3
=
\frac{\left(1+i\right) e^{-\frac{i \left(\sqrt{2}\, \e+1\right)}{\e}}}
{12\,\sqrt[4]{8\,\e}\,\sqrt{k}}
&&
\!\!\!
\left\{
\sqrt{\e} \left[\left(\sqrt{2}+2\, i\right) e^{2\, i\,\sqrt{2}}
-\left(\sqrt{2}-2\, i\right) e^{2\, i/\e}\right] D_{\frac{1}{2}}\left(\frac{1-i}{\sqrt{\e}}\right)
\right.
\nonumber
\\
&&
\left.
+(1-i) \left(\sqrt{2}-2\,i\right) e^{2\, i/\e} D_{-\frac{1}{2}}\left(\frac{1-i}{\sqrt{\e}}\right)
\right\}
\ .
\eeqa
The power spectrum is finally obtained to be  
 \beq
P_{\rm CJ}^{(3)}
\equiv
\gamma^{(3)}\,P_{\rm BD}
=
2\, |A_3|^2\, P_{\rm BD}
\ . 
\eeq
In fig.~\ref{gammas-compared}, we plotted the factor $\gamma^{(3)}$ as a function of $\e$.
One can see that the factor $\gamma^{(3)}$ oscillates around $4/3$ for small values of $\e$,
with an amplitude roughly proportional to $\e$.
This oscillatory behaviour continues for larger values of $\e$, up to the point of validity of the above computation. 
Note that when we are matching two oscillatory solutions at high frequencies we expect the error to oscialite as well.  
As we see the exact result does not show any oscillations, but only a suppression
proportional to $-\e^2$ close to $\e\approx 0$. 
The large offset of $1/3$ from the exact result is reminiscent of the discrepancy observed in the previous section 
from gluing the modes at the horizon crossing point, $x=-\sqrt{2}$.
In fact, the offset can be removed if the regions~II and~III are combined.
The improved solutions in this unified region, 
\eqref{excitedmode},
can then be glued to the solution in region~I.
As we see in fig.~\ref{gammas-compared}, the modulation factor $\gamma^{(2)}$ obtained using
just two region does not show any offset from the exact result anymore.
This result still displays an oscillatory feature for small values of the distortion parameter $\e$,
which is again due to the inadequacy of this technique in the UV.  
\par

\begin{figure}[t]
\begin{center}
\includegraphics[angle=0, scale=1.3]{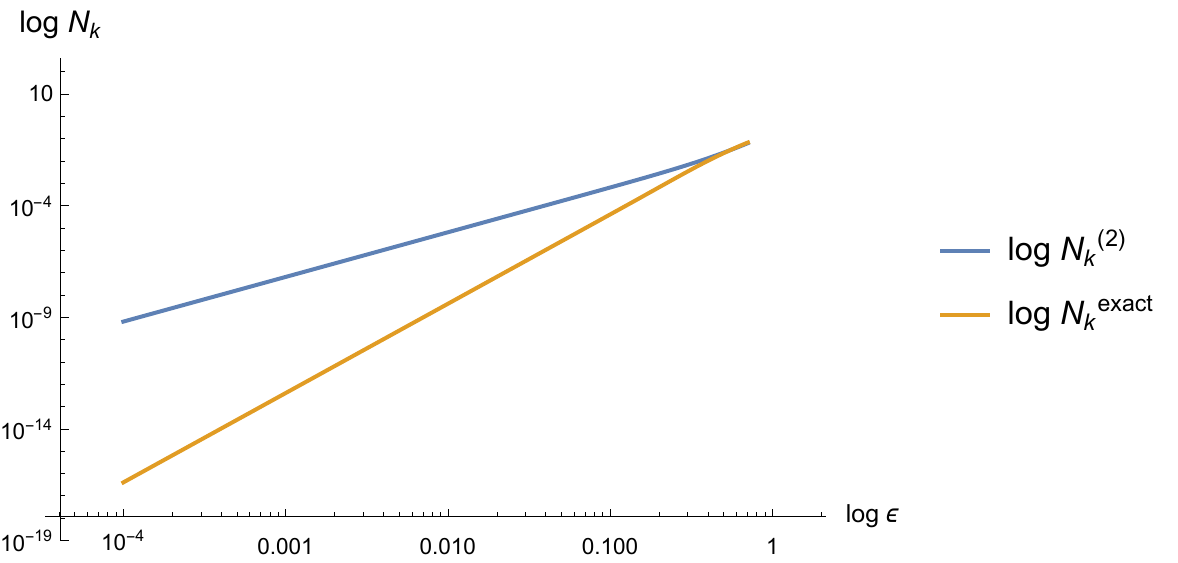}
\caption{Number of particles in the corresponding excited states obtained via gluing is compared to the exact one as function of $\e$ . The exact number of particles goes to zero faster by an extra factor of $\e^2$. }\label{Nk-e}
\end{center}
\end{figure}

The number of particles obtained from the gluing technique with two-regions 
is 
\beqa
N_k^{\rm (2)}(\e)
&=&
\frac{\pi} {4 \,\sqrt{2}\,\epsilon ^{3/2}}
\left\{
D_{-\frac{1}{2}}\!\left(\frac{1-i}{\sqrt{\epsilon }}\right)
\left[
H_{\frac{1}{2}}^{(1)}(\epsilon^{-1})
-(\epsilon -i)\,H_{\frac{3}{2}}^{(1)}(\epsilon^{-1})
\right]
\right.
\nonumber
\\
&&
\left.
+(1-i)\,\sqrt{\epsilon }\,D_{\frac{1}{2}}\!\left(\frac{1-i}{\sqrt{\epsilon }}\right)
H_{\frac{3}{2}}^{(1)}(\epsilon^{-1 })
\right\}
\times
\left\{
(1+i)\,\sqrt{\epsilon } \,D_{\frac{1}{2}}\!\left(\frac{1+i}{\sqrt{\epsilon }}\right)
H_{\frac{3}{2}}^{(2)}(\epsilon^{-1 })
\right.
\nonumber
\\
&&
\left.
+D_{-\frac{1}{2}}\left(\frac{1+i}{\sqrt{\epsilon }}\right)
\left[H_{\frac{1}{2}}^{(2)}(\epsilon^{-1 })
-(\epsilon +i)\,H_{\frac{3}{2}}^{(2)}(\epsilon^{-1 })
\right]
\right\}
\ ,
\eeqa
which behaves like $\e^2/16$ for small $\e$, as opposed to $25\, \e^4/64$ for the exact solution.
In fig.~\ref{Nk-e},
we have plotted both the number of particles obtained by the gluing method and the one
from the exact mode functions. It displays it clearly that the number of particles for exact solution goes to zero faster than
what the gluing method predicts for small $\e$, .
\section{Sixth order polynomial dispersion relation}
\label{S6o}
We are now going to take one step further and investigate mode equations of the form \eqref{disp6}, where the dispersion relation is governed by $\omega^2\propto k^6$ in the infinite past. In particular we consider the cases that perturbation modes start from adiabatic vacuum, then go through a non-adiabatic phase where group velocity becomes negative and finally re-emerge as super-excited states when $\omega^2\simeq k^2$.
For convenience we have set the sound speed in \eqref{disp6} to unity~\footnote{If $c_s \neq1$, one can still make it to one in mode equation for the canonical variable $u$, by changing the conformal time $d\tau \to c_s d\tau $. When evaluating the power spectrum for field $\phi$ or curvature perturbations that factor will reemerge.}. Implementing this dispersion for the the canonicalized variable $u$, mode equation takes the following form
\beq
\label{disp6-2}
u_k''+\left( \beta_0\, x^4-\alpha_0\, x^2+ 1 -\frac{2}{x^2} \right) u_k
=
0
\ ,
\eeq
where $x\equiv k\tau$ as before, $\beta_0\equiv \bar{\beta} H^4$ and $\alpha_0\equiv \bar{\alpha} H^2$. As we will discuss in section~\ref{Seft}, these dispersion relations can be realized, in the effective field theory of inflation~\cite{Cheung:2007st}, from terms like
$\nabla_{\mu} \delta K^{\nu\gamma}\,\nabla^ {\mu} \delta K_{\nu\gamma}$,
$(\nabla_{\mu} \delta K^\nu {}_\nu)^2$, $\nabla^ {\mu}\delta K_{\nu\mu}\nabla^ {\nu}\delta K_{\sigma}^{\sigma}$ and
$\nabla_{\mu} \delta K^\mu {}_\nu\, \nabla_{\gamma} \delta K^\gamma {}^\nu $.
A similar dispersion relation has also come up in the study of trans-Planckian signatures in
inflation~\cite{Martin:2003kp,Joras:2008ck}. We can constraint the dispersion relation \eqref{disp6} to be nonnegative (non-tachyonic) at all sub Hubble scales by setting \cite{Joras:2008ck},
\beq
z\equiv \frac{\beta_0}{\alpha_0^2}\geq \frac{1}{4}\,.
\eeq
To be on the conservative side, we also assume that the modes get lighter than the Hubble scale only once during inflation,
which corresponds to the time of horizon-crossing, {\rm i.e} $\omega^2=2H^2$ has only one solution which is around $k \sim 1/aH$ . This is different from the study of  Ref.~\cite{Joras:2008ck},
where it was assumed that there are three turning points corresponding to three solutions of the equation
$\omega^2-2\,H^2=0$.
For 
\beq
z>\frac{1}{3},
\eeq
this equation automatically has only one solution.  For values of $\az$ and $\bz$ such that
\beq
\frac{1}{4}\leq z \leq \frac{1}{3}\,,
\eeq
having only one solution imposes one of the following conditions
\beqa \label{az-range}
\az &\leq&  \frac{9 z-2-2 (1-3 z)^{3/2}}{54 z^2} \nonumber \\ 
\text{or} \quad \az &\geq& \frac{9 z-2+2 (1-3 z)^{3/2}}{54 z^2}\,.
\eeqa
\par
Again, we will first try to estimate the particle number density using the gluing approach. 
We decompose the domain of $x\equiv k\tau\in(-\infty,0]$ into the intervals
\beqa
-\infty<x\lesssim x_1(\az,\bz)
\qquad
&&{\rm region~I}
\\
x_1(\az,\bz)<x<0
\qquad
\qquad
&&{\rm region~II}
\eeqa
where $x_1 (\az,\bz)$ is the transition point below which the higher order corrections to the dispersion relation
can be neglected and the mode equation becomes the stabdard Lorentzian dispersion relation.
In general $x=x_1(\az,\bz)$ can be determined by solving the equation
\beq
\az x^4-\bz x^2=1-\frac{2}{x^2}
\ ,
\eeq
and is a complicated function of $\az$ and $\bz$, but we note that for the cases of our interest $\bz\sim \az^2$,
and one has  
\beq
x_1(\az,\bz)\approx -\frac{2}{\sqrt{\az}}\,.
\eeq
Interestingly, Maple can find a compact solution in terms of Heun~T functions  to the mode equation if 
the term ${2}/{x^2}$ was dropped. Therefore this solution is applicable to region I.  After normalizing it according to the Wronskian condition~\eqref{Wronskian},
and imposing  adiabatic vacuum initial condition, we obtain 
\beq
u_{k\,\rm I}(x)
=
\frac{\bz^{1/4}}{k^{1/2}\,\alpha_0^{1/2}} {\rm HeunT}(\mathscr{ A}, 0, \mathscr{B}, -\mathscr{C}~x)\, \exp(-y)
\ ,
\eeq 
where
\beqa
\mathscr{ A}
&=&
\frac{1}{32}\,{\frac {{\az}^{2} 18^{1/3}  \left( i\sqrt {3}+1
 \right) ^{2}}{{\bz}^{4/3}}}
\nonumber
\\
\mathscr{ B}
&=&
-\frac{1}{4}\,{\frac {\az\,\sqrt [3]{12} \left( i\sqrt {3}+1 \right) }{{
\bz}^{2/3}}}
\nonumber
\\
\mathscr{ C}
&=&
-\frac{1}{3}\,{3}^{2/3}\sqrt [3]{2}\sqrt [6]{\bz}\sqrt [6]{-1}\ ,
\eeqa
and
\beq
y
=
\frac{1}{12}\,{\frac {x\, \left( -4\,i\bz\,{x}^{2}+3\,\sqrt {3} \left( -1 \right) ^{2/3}\az+3\,\sqrt [6]{-1}\az \right) }{\sqrt {\bz}}}\ .
\eeq
This solution has to be matched to the general solution in region~II, which is nothing other than a linear combination of Hankel functions~\eqref{excitedmode},
\beq
\label{Hankel-mode}
u_{k\,\rm II}
=
\frac{\sqrt{-x\pi}}{2\sqrt{k}}\left[\xi \,H_{3/2}^{(1)}(-x)+\rho \,H_{3/2}^{(2)}(-x)\right]
\ .
\eeq
The solutions should be glued following the same prescription as~\eqref{gluing-prscription1} and ~\eqref{gluing-prscription2} at the point
$x_g=x_1(\az,\bz)$. 
\begin{figure}[t]
\begin{center}
\includegraphics[angle=0, width=0.45\textwidth]{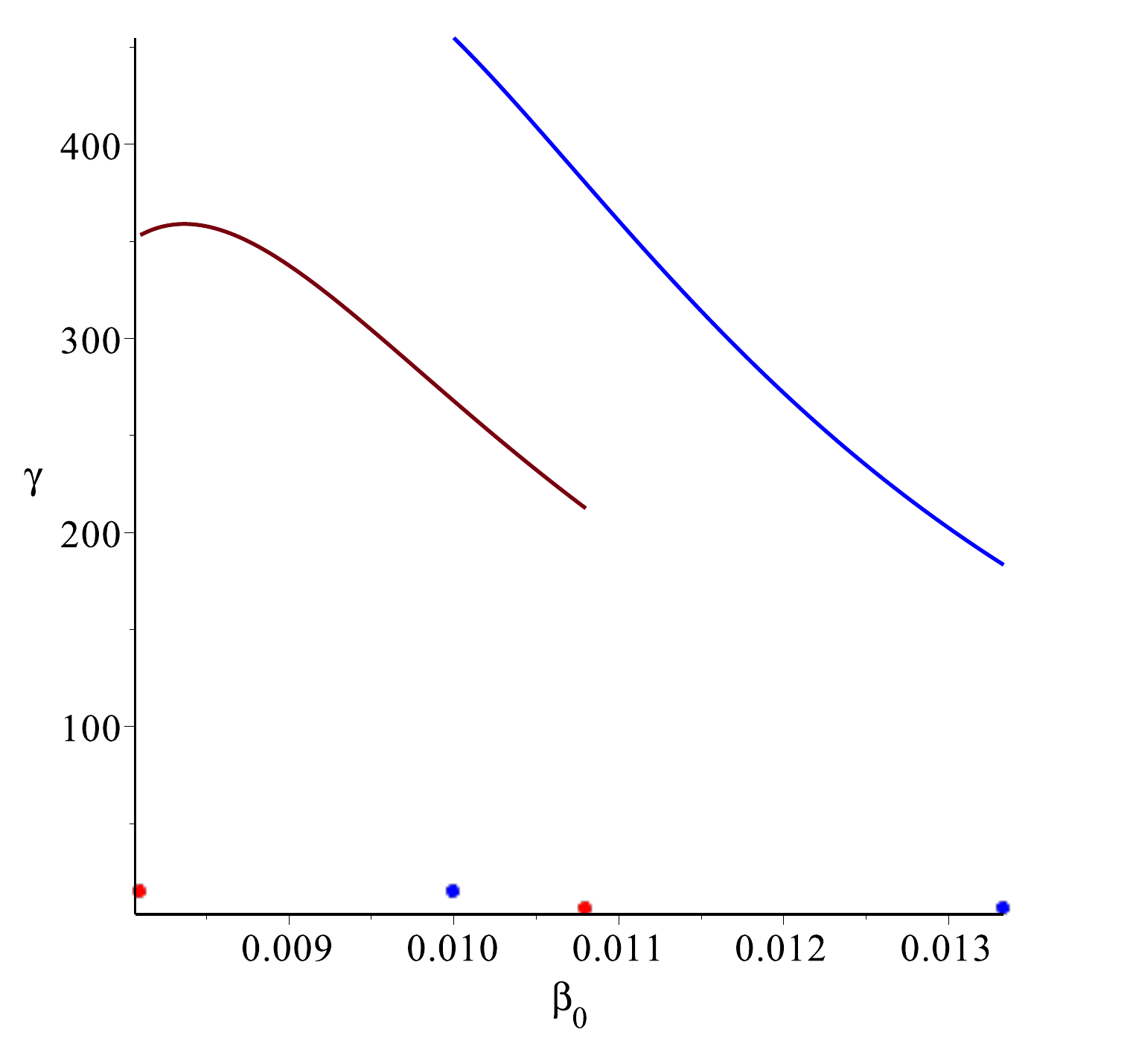}
\includegraphics[angle=0, width=0.45\textwidth]{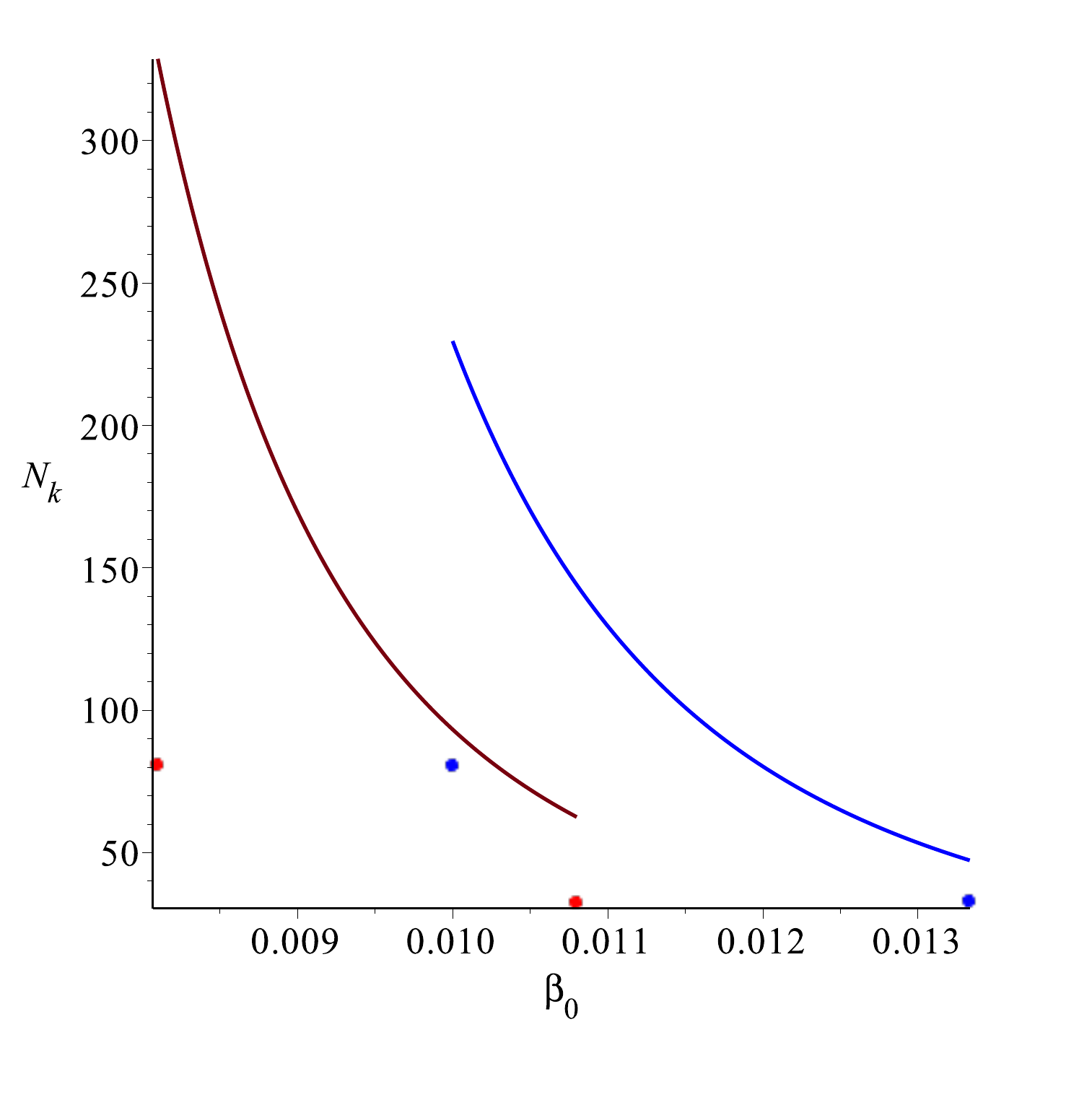}
\caption{Power spectrum enhancement (left panel) and occupation number (right panel) obtained with gluing method (red and blue curves) compared to full numerical result (dots). For few values of $\alpha_0$ and $\beta_0$, we have computed the correction to the power spectrum and the number of particles in the corresponding excited state numerically. The difference between these dots and the corresponding point on the curves shows that the gluing method could have a large error in estimating these parameters. The blue and red curves (dots), correspond to $\alpha_0=0.2$ and $\alpha_0=0.18$, respectively.  }
\label{gammaNk}
\end{center}
\end{figure}
\par
For $\az=0.2$ and 
\beq
\frac{\az^2}{4}\leq \bz\leq \frac{\az^2}{3}
\eeq 
Eqs. \eqref{az-range} is satisfied. For larger ratios of ${\bz}/{\az^2}$, although one can solve for the Bogolyubov coefficients implicitly in terms of Heun~T and Heun~T~Prime functions, Maple software cannot evaluate the values of these parameters numerically.
Apparently Maple employs a series expansion for the Heun~T~Prime function which does
not converge for such range of parameters. For $\az=0.2$,  the enhancement factor for power spectrum is in the following range 
\beq
183.35\leq \gamma\leq 454.89,
\eeq
which is quite large. The range for number of particles in the excited state, $N_k$, is
\beq
47.31\leq N_k \leq 229.63
\eeq
In Fig.~\ref{gammaNk}, we have plotted $\gamma$ and $N_k$ for $\az=0.2$ and $\az=0.18$, in the range $\frac{1}{4}\leq z\leq \frac{1}{3}$. Such values of $\az$ satisfy the constraints \eqref{az-range} and thus there is only one turning point corresponding to the horizon crossing. As we discussed in the last section,
the gluing technique can become unreliable in obtaining the enhancement parameter and the number of particle. Thus, one may wonder to what extent the above results can be trusted. The following derivation shows that in fact, compared to the numerical result, gluing method significantly overestimates the occupation number of excited states and the enhancement factor. 

\begin{figure}[t]
\begin{center}
\includegraphics[angle=0, scale=1.3]{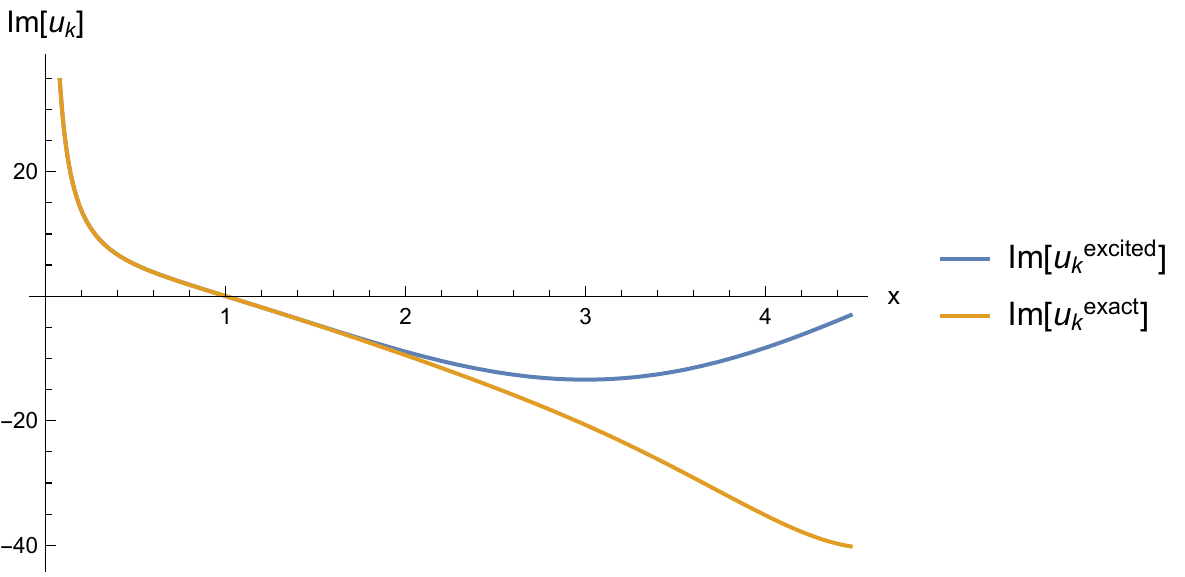}\\
\includegraphics[angle=0, scale=1.3]{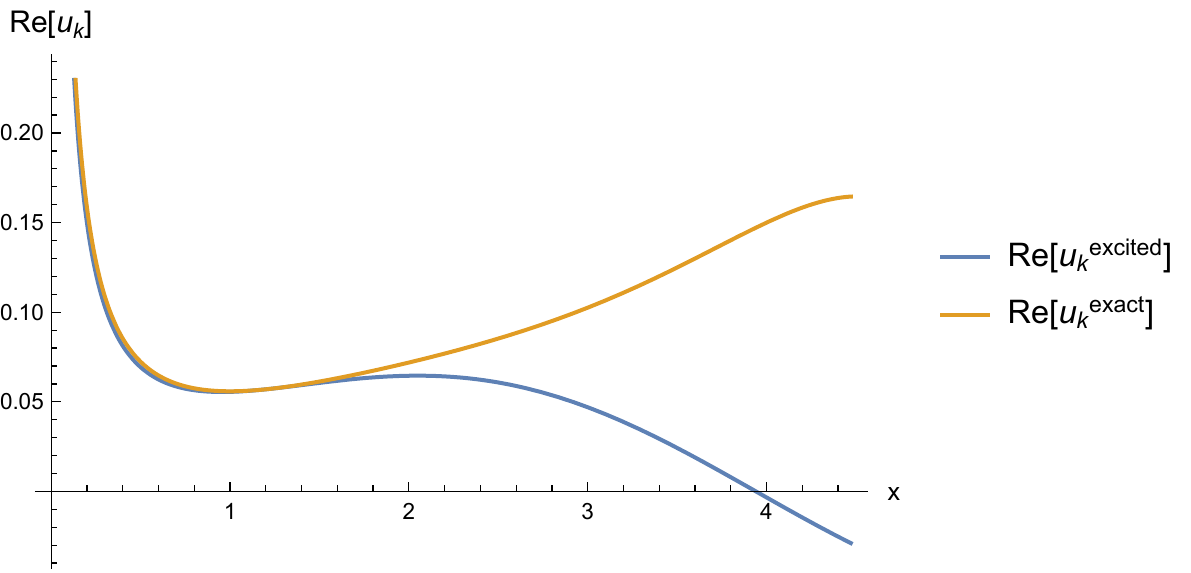}
\caption{Imaginary and real parts of the implicit mode function is compared with the excited state
for the choice of $a$ that results in the coalescence of the modes after the collision point.}
\label{uks-comparison}
\end{center}
\end{figure}
\par
Although one cannot find the exact power spectrum parametrically as a function of $\az$ and $\bz$,
we can still compute the power spectrum numerically for fixed values of these parameters.  Starting from the normalized positive frequency WKB mode in the infinite past,
\beq
\label{initcond2}
u_k(x\to-\infty)
\simeq
\frac{1}{2} \left(-\frac{\pi }{3} x\right)
H_{\frac{1}{6}}^{(1)}\left(- \frac{\sqrt{\beta_0}}{3}\,x^3\right)
\ ,
\eeq
the solution is then evolved numerically. 
Table~\ref{table}, lists the values of the enhancement factor and other variables evaluated  numerically,
for given pairs of $(\az,\bz)$.
The largest enhancement for the power spectrum occurs when
$\alpha_0\simeq 0.2 $ and $\beta_0\simeq\alpha_0^2/4$ and, for illustration purposes,
we focus on these particular values of parameters.
The enhancement factor over the Bunch-Davies result for the power spectrum
is about $\gamma=14.738$. This is much smaller than the enhancement factor, $\gamma=454.89$, obtained through gluing. Contrary to the adiabatic case, the gluing technique fails by a huge amount with respect to the numerical result.

Although Mathematica does not recognise any compact explicit solution for the mode equation~\eqref{disp6-2}
inside the horizon, it can come up with implicit solutions for the complete mode equation, which 
formally read as
\beq
\label{implicitsol}
u_k(x)
=
c_1\, u_k^{(1)}(x)+c_2\, u_k^{(2)}(x)
\ ,
\eeq
where $c_1$ and $c_2$ satisfy 
\beq
c_1 \,\bar{c_2}-c_2\, \bar{c_1}
=
i
\ .
\eeq
The solutions $u_k^{(i)}$, with $i=1,2$, are normalized so that for one of them $u_k^{(i)}(1)=1$
and $u_k^{(i)'}(1)=0$ and for the other one it is the other way around\footnote{It turns out that these implicit solutions are much easier
to evaluate if their argument is real and positive.
Noting that the differential equation is even under $x\rightarrow -x$, we will work in the domain
$(0,\infty)$, instead of $(-\infty,0)$.}.
The most general solutions to the Wronskian condition with real  $c_1$ is 
\beqa
c_1
&=&
\frac{1}{\sqrt{2 }\,s }
\,
\\
c_2
&=&
-i\frac{s}{\sqrt{2}}
\ ,
\eeqa
where $s\in \mathbb{R}$.
We determine the parameter $s$, by requiring that the power spectrum obtained from the
modes~\eqref{implicitsol} matches the one obtained by numerically evolving the initial
condition~\eqref{initcond2}.
There are usually four different solutions, two by two negative of each other.
For all these values of $s$, one can compute the corresponding implicit mode function
and we choose the one which leads to almost the same value of the mode function obtained
by integrating numerically the mode equation.
In principle, one can then follow the steps taken in the previous section to determine the number
of particles in the corresponding state.
However, while the amplitude of fluctuations, which is proportional to $f(x)\equiv x \,u_k(x)$,
approaches a constant value outside the horizon, the way Mathematica numerically solves the equation, the error accumulates in higher derivative terms such as  $f'''$, and make them diverges as $x\to 0$.  
What we do instead is the following, we expect the general solution of ~\eqref{disp6-2} to merge quite well to Henkel functions even before Hubble crossing. Therefore, we glue the mode function~\eqref{implicitsol} to a linear 
combination of Hankel functions at an {\it arbitrary\/} point $x_g$.
We next demand the corresponding Bogolyubov coefficients $\xi(x_g)$ and $\rho(x_g)$ yield the numerically
computed value for the power spectrum.  We then test our solution to make sure our assumption for gluing at $x_g$ is justified. 
For $\az=0.2$ and $\bz=\az^2/4$, we have plotted both the real and imaginary parts of the implicit
numerical solution and the excited one obtained though matching the linear combination of Hankel functions (see fig.~\ref{uks-comparison}). In this case $x_g=1.2145$, and as the figure displays these functions merge very well at that point. The values
of the Bogolyubov coefficients estimated at this point are
\beq
\xi(x_g)
=1.95519 - 8.80935\,i
\,
\qquad
\rho(x_g)=-1.88359 - 8.7681\,i
\ ,
\eeq
which suggests that the particle number density in the corresponding excited state is $N_k=80.4275$.
This number is quite large and suitable to serve as a highly excited state above the Bunch-Davies
vacuum. In the language of \cite{Ashoorioon:2013eia}, this corresponds to $\chi_S\simeq 2.19$, which is enough to serve our interests.

In table~\ref{table}, we also list the corresponding values of the Bogoliubiov coefficients 
that yields the correct enhancement factor to the power spectrum. 
\begin{table}[t]
\begin{center}
\begin{tabular}{ |c|c|c|c|c|c|} 
 \hline
 $\az$ & $\bz$ & $\gamma$ & $\rho$ & $\xi$ &$N_k$\\ \hline
 $0.1$&$\az^2/3$  & $2.62652$   &$-0.727078 - 3.65736~i$ &$0.890536 - 3.75658~i$&$13.9049$\\ \hline
  $0.18$&$\az^2/4$  & $14.7378$   &$1.88601 + 8.77648~i$ &$-1.95745 + 8.81771~i$&$80.5836$\\ \hline
   $0.18$&$\az^2/3$  & $5.95909$   &$-1.16428 - 5.55538~i$ &$1.27598 - 5.62047~i $&$32.2178$\\ \hline
    $0.19$&$\az^2/4$  & $14.7739$   &$-1.88609 - 8.77973 ~i$ &$1.95736 - 8.82098~i$&$80.641$\\ \hline
    $0.19$&$\az^2/3$  & $5.95909$   &$-1.16407 - 5.55488~i$ &$1.27619 - 5.61989~i $&$32.2118$\\ \hline
    $0.2$&$\az^2/4$  & $14.738$   &$-1.88359 - 8.7681~i$ &$-1.95519 + 8.80935 ~i$&$80.4275$\\ \hline
     $0.2$&$\az^2/3$  & $6.06496$   &$-1.17517 - 5.60415 ~i$ &$1.2867 - 5.6685 ~i$&$32.7875$\\ \hline
     $0.3$&$\az^2/4$  & $12.838$   &$-3.66837 + 1.44445 ~i$ &$1.83104 - 8.21655 ~i$&$69.8644$\\ \hline
     $0.3$&$\az^2/3$  & $6.06576$   &$-1.17328 - 5.59953 ~i$ &$1.28877 - 5.66307 ~i$&$32.7313$\\ \hline
     $0.4$&$\az^2/4$  & $10.6292$   &$-1.58475 - 7.42534 ~i$ &$1.67516 - 7.47268 ~i$&$57.647$\\ \hline
     $0.4$&$\az^2/3$  & $5.44107$   &$-1.10289 - 5.29483 ~i$ &$-4.44333 + 1.44564 ~i$&$29.2516$\\ \hline
     $0.5$&$\az^2/4$  & $8.87117$   &$-1.43795 - 6.77284 ~i$ &$1.54006 - 6.82402 ~i$&$47.939$\\ \hline
       $0.5$&$\az^2/3$  & $4.79253$   &$1.16308 - 5.03005 ~i$ &$-1.02498 - 4.9602 ~i$&$25.6542$\\ \hline
\end{tabular}
\end{center}
\caption{Modulation factor $\gamma$, Bogolyubov coefficients $\xi$ and $\rho$, and particle number density
$N_k$ for the sixth order dispersion relation.}
\label{table}
\end{table}
\par 

We note in passing that it is also possible to produce the so called "calm excited states".~\cite{Ashoorioon:2010xg}, with this type of dispersion relations. One can cook up values for $\az$ and $\bz$ that lead to excited states but do not modify the
power spectrum at all.
For example, with
\beqa
\az
&=&
0.0101982725
\ ,
\\
\bz
&=
&\frac{\az^2}{3}
\ ,
\eeqa
one obtains an excited state with Bogolyubov coefficients
\beqa
\rho
&=&
-0.320196 - 2.16841\,i
\ ,
\\
\xi
&=&
0.669079 - 2.31449\,i
\ ,
\eeqa
and particle number density $N_k\simeq 4.80453$. Nonetheless, the impact of such an excited state on the two-point function is negligible. Therefore, 
one cannot say conclusively if the mode has originated from an excited state just by examining
the power spectrum.
\section{Higher order dispersion relations in the EFT of inflation}
\label{Seft}
One way to understand the appearance of higher order terms in dispersion relations such as those discussed in the last section, is by applying the approach of Effective Field Theory (EFT) of inflation \cite{Cheung:2007st}. This approach allows one to write down the most generic action for models with one scalar field $\phi$ on a quasi de Sitter background. 
The idea is that even though the action is invariant under all diffeomorphisms (diffs), the solution for $\phi(x^\mu)$ provides a preferred time slicing. Restricting time hypersurfaces to be constant $\phi$ surfaces (the unitary gauge)\footnote{Note that this choice doesn't completely fix the gauge. There is still a remaining gauge freedom in scalar metric perturbations due to spatial diff $x^i\rightarrow x^i+\delta^{ij}\partial_j \xi$.}, then action only has to satisfy 3d spatial diff invariance. The most general spatial diff invariant action, around FRW metric, is given by \cite{Cheung:2007st}
\beq 
S
=
\int d^4x\, \sqrt{-g}
\left\{
M_{\rm Pl}^2
\left[
\frac12\, R
+\dot H\,  g^{00}
-\left(3\, H^2 +\dot H\right)
\right]
+L_n
\right\}
\ ,
\label{genspacei}
\eeq
where
\beq
L_n
=
\sum_{m\geq 2}^n\,
F^{(m)}(g^{00}+1,\delta K_{\mu\nu}, \delta R_{\mu\nu\rho\sigma};\nabla_\mu;t)
\ .
\eeq
Here, $F^{(m)}$ represent functions that are of order $(m)$ in metric perturbations. 
This is because in addition to 4d diff invariant metric terms, now $g^{00}$ and pure functions of time $f(t)$ are also scalars under 3d spatial diffs. We can also include terms that have extrinsic curvature $K_{\mu\nu}$ of the constant time hypersurfaces as it is a tensor under spatial diffs. Any additional 3d diff invariant object such as 3d covariant derivative can be rewritten in terms of the terms already included. Furthermore, one expects to recover FRW solution at the zeroth order in perturbation which has been used to fix the terms inside the square bracket. 

Now that we have the generic action in unitary gauge, we can relax the gauge condition by time transformation $t\to t+\xi^0(x^\mu)$. Since the action is no more restricted to a particular time slicing, we need to restore time diff invariance again. This can be achieved by promoting $\xi_0(x^\mu)$ to a field $-\pi(x^\mu)$ and requiring that it shifts as $\pi(x^\mu)\rightarrow \pi(x^\mu)-\xi_0(x^\mu)$ under the time diffs. 
Note that when we perform $t\to t+\xi^0(x^\mu)$, we no longer expect the perturbations in $\phi$ to be zero.  In fact, by introducing the goldstone boson we are representing these perturbations since  $\delta\phi=-\dot{\phi}_0\xi^0$ and therefore $\pi=\delta\phi/\dot{\phi}_0$. 

In general this action is going to be very complicated. However, the advantage of using this approach is that when the generalized slow roll approximation is valid, one can ignore all the metric perturbations in the action. Basically, if the time dependence of the coefficients in unitary gauge are much slower than Hubble time, then the decoupling limit of this theory from gravity, is larger than Hubble scale. This makes it possible to calculate power spectrum neglecting metric perturbations in the action. 

For instance, if we implement this procedure to action \eqref{genspacei}, assuming $L_n=0$, we obtain
\beq
S_\pi= -M_{pl}^2\int d^4x\, \sqrt{-g} \dot{H}\left (\dot{\pi}^2-{(\partial \pi)^2\over a^2}\right). 
\eeq
One can show through gauge transformation, that $\pi$ is related to the conserved quantity $\zeta$, through $\zeta=-H\pi$. Which after substituting in above action reduces to standard slow roll inflationary action for $\zeta$. However, we are interested at modification of dispersion relation beyond standard slow roll models. This task can be achieved by turning on the coefficients in $L_n$. Keeping the terms that can contribute to the quadratic action of $\pi$, at most to sixth order in gradients expansion and are even under time reversal, $L_n$ reduces to 
\begin{eqnarray}
\label{eq:actiontad}
L_{n}
&=&
\frac{M_2^4}{2!}\,(g^{00}+1)^2
-\frac{\bar M_2^2}{2}\, (\delta K^\mu_{\ \mu})^2
-\frac{\bar M_3^2}{2}\, \delta K^\mu_{\ \nu}\,\delta K^\nu_{\ \mu}
\nonumber
\\
&&
-\frac{\delta_1}{2}\,(\nabla_{\mu} \delta K^{\nu\gamma})(\nabla^ {\mu} \delta K_{\nu\gamma})
-\frac{\delta_2}{2}\,(\nabla_{\mu} \delta K^\nu_{\ \nu})^2 
-\frac{\delta_3}{2} \,(\nabla_{\mu} \delta K^\mu_{\ \nu})(\nabla_{\gamma} \delta K^{\gamma\nu})\nonumber
\\
&&
-\frac{\delta_4}{2} \,\nabla^ {\mu}\delta K_{\nu\mu}\nabla^ {\nu}\delta K_{\sigma}^{\sigma}
\ .
\label{genv}
\end{eqnarray}
$M_i$ and $\delta_i$ are free time dependent coefficients and the sign of each term is also a priori free. 
As pointed out in \cite{Cheung:2007st,Bartolo:2010bj}, please also see \cite{Gwyn:2012mw}, after proper canonicalization terms in the first line of eq.~\eqref{eq:actiontad}
can lead to an equation of motion with quartic corrections to the dispersion relation,
\beq
u''+\left( \gamma_0k^2+\alpha_0 \,k^4\, \tau^2 -\frac{2}{\tau^2}\right)u
=0
\ ,
\eeq
where $\alpha_0$ and $\gamma_0$ are functions of the $M_i$ and $\bar{M_i}$ parameters.
Noting that 
\beq
\delta K_{ij}
\supset
(\partial_i\partial_j \pi+\pa_ig_{0j})
\ ,
\eeq
we expect the correction in the second line of equation \eqref{genv} give rise to 
\beq
u''
+\left( \gamma_0 k^2+\alpha_0 k^4 \tau^2 +\beta_0 k^6 \tau^4 -\frac{2}{\tau^2}\right) u
=0
.
\eeq
Evaluating the action explicitly for $\pi$ in Fourier space, we find that in $\dot{H}\rightarrow 0$ and decoupling limit, the second line of \eqref{eq:actiontad} produces these quadratic  terms

\begin{eqnarray}
\mathcal{L}_{n}^{(2nd)}&=&-\frac{1}{2} \delta_1 \left(\frac{k^6 \pi ^2}{a^6}-\frac{3 H^2 k^4 \pi ^2}{a^4}-\frac{k^4 \dot{\pi}^2}{a^4}+\frac{4 H^4 k^2 \pi ^2}{a^2}-6 H^4 \dot{\pi}^2-3 H^2 \ddot{\pi}^2\right)\nonumber\\&&
-\frac{1}{2} \delta_2 \left(\frac{k^6 \pi ^2}{a^6}+\frac{H^2 k^4 \pi ^2}{a^4}-\frac{k^4 \dot{\pi}^2}{a^4}+\frac{6 H^4 k^2 \pi ^2}{a^2}-9 H^2 \ddot{\pi}^2\right)\nonumber\\&&
-\frac{1}{2} \delta_3 \left(\frac{k^6 \pi ^2}{a^6}+\frac{3 H^2 k^4 \pi ^2}{a^4}+\frac{ H^2 k^2 {\dot{\pi}} ^2}{a^2}-9 H^4 \dot{\pi}^2\right)\nonumber\\&& -\frac{1}{2}\delta_4 \left(\frac{k^6 \pi ^2}{a^6}+\frac{ H^2 k^4 \pi ^2}{2 a^4}+\frac{9 H^4 k^2 \pi ^2}{2 a^2}+\frac{3 H^2 k^2 \dot{\pi}^2}{a^2}+\frac{27}{2} H^4 \dot{\pi}^2\right)\,.\nonumber\\
\end{eqnarray}
As we expected the ~$k^6\pi^2$ do appear in $L_n$ corrections. $-\frac{\delta_1}{2}(\nabla_{\mu} \delta K^{\nu\gamma})(\nabla^{\mu} \delta K_{\nu\gamma})$ and $\frac{\delta_2}{2}(\nabla_{\mu} \delta K^\nu_{\ \nu})^2 $ terms also produce  $\ddot{\pi}^2$. According to Ostrogradski theorem \cite{Ostrogradski}, higher time derivatives usually lead to ghost instabilities. To avoid this, we set $\delta_1=\delta_2=0$. In addition, the presence of terms like $k^4\dot{\pi}^2$ or $k^2 \dot{\pi}^2$ will modify the dispersion relation in the UV back to $\omega^2 \propto k^4$ and $\omega^2 \propto k^2$ again. Hence, we set $\delta_3=-3\delta_4$ to avoid this modification\footnote{In the previous copy of this paper, we have evaluated  $\delta K_{\gamma\nu}$ by directly subtracting FRW extrinsic curvature of constant time surface($K^{(0)}_{ij} = a^2 H \delta_{ij}$, $K^{(0)}_{00}=K^{(0)}_{i0}=0$) from $K_{\mu\nu}$. However, with that definition $\delta K_{\gamma\nu}$ is not a proper tensor, so we redid our calculation using $K^{(0)}_{\mu \nu} = H h_{\mu \nu}$, where $h_{\mu \nu}$ is the induced metric of constant time surfaces \cite{Cheung:2007st}. This change modified some terms and coefficients, and added $k^2\dot \pi^2$ terms to both $\delta_3$ and $\delta_4$ terms.}. The remaining terms can contribute to the quadratic, quartic and sextic corrections to the dispersion relation. In general the coefficients of these correction can be either positive or negative. In this particular paper we focused on the example where the quadratic and sixth order coefficients, appear with a positive sign but the coefficient of quartic correction, $\az$, is negative. That particular choice and avoiding superluminal propagation, are in principle possible, by tuning the parameters \footnote{While with the the higher order corrections to the dispersion relation, one may face UV superluminality, one can still avoid superluminal propagation the IR mode by requiring $c_s\equiv \sqrt{\gamma_0}\ll 1$.}. We leave a more detailed exploration of the parameter space for $\delta_3$ and other coefficients, that result in super excited or calm states for future work. 

Even though such dispersion relations are in principle easy to produce within the effective field theory of inflation, it has been argued \cite{Cheung:2007st} that invoking dispersion relations $\omega^2 \propto k^{2n} $ with $n\geq 3$ undermines the concept of effective field theory itself. This is because for $n\geq 3$ the interaction terms scale with negative power of energy and therefore become strongly coupled at low energies.  That comes from the fact that the scaling dimension of $\pi$ will not be the same as the one for the Lorentzian dispersion relation. This implies that in the high momenta regime where we are invoking $\omega^2\propto k^6$, theory has an IR cut off, below which it becomes strongly coupled. However, note that as universe expands the quartic and Lorentzian corrections to dispersion relation become more important and eventually take over. Given the different parameters that can contribute to kinetic terms, we think it is plausible to tune the IR cut off for that regime to be below the scale where dispersion relation transitions to $ \omega^2 \propto k^2 - \bar{\alpha} k^{4}$. One may argue that this makes the theory highly fine tuned at UV. That may very well be the case, but given that we don't know what the UV complete theory is, it is hard to say what is natural and what terms are protected under the fundamental symmetries. For example Horava's  proposal for a UV completion of gravity \cite{Horava:2009uw} naturally predicts dispersion relation of the form $\omega \propto k^{3}$ for graviton, based on Lifshitz symmetry. 
In the end, we also expect that the overall bounds on the non-gaussianity be satisfied when speed of sound is close to one \cite{Ashoorioon:2013eia}. The non-gaussianity parameter, $f_{\rm NL}$, remains finite for super-excited states, since it is defined as  bispectrum divided by the power spectrum squared. Even though three point functions for such 
super-excited states may get enhanced \cite{Shukla:2016bnu}, as we showed here the power spectrum is enhanced too, so the net effect on $f_{\rm NL}$ may still remain small. 

\section{Conclusion}

We constructed super-excited initial conditions around the new physics hypersurface, with the help of modified dispersion relations. In this peculiar type of dispersion relations, there is an intermediate phase where dispersion relation has a negative slope and results in increasing energy of the modes while their wavelength expands. Still the energy of the modes remain larger than Hubble parameter while they are sub horizon. Even though the modes start from adiabatic vacuum in infinite past, due to this intervening  phase, there will be a substantial amount of particle production if the evolved state is mapped to standard Lorentzian modes. Such excited initial states can have interesting phenomenology and it is interesting to know if they are motivated from a more fundamental perspective. We demonstrated that these states can possibly appear within the effective field theory of inflation where the existence of higher dimensional operators naturally leads to the modified dispersion relation. It is interesting to further investigate the bispectrum of such dispersion relations and work out the allowed region of parameters explicitly. The possibility of sixth order and higher dispersion relations were dismissed in the earlier investigations of effective field theory of inflation noting that in the IR they will be strongly decoupled and invalidate the effective field theory description. The argument is based on dimensional energy scaling argument that evaluates the dimension of the goldstone boson scalar field based on dispersion relation. However, in an expanding background, where the higher dimensional terms get redshifted as the universe expands, such dimensional arguments does not seem to be strong. Still, at low energies, the dispersion relation reduces to the Lorentzian one and, a priori, it is not clear that evolution of the mode during the time the physical momentum of the mode is in the new physics region, can lead to effect that makes the effective field theory invalid. In addition, the three point function needs to be compared with the two point function, which as we showed in this article, is by itself large. This analysis is something we plan to return to in future.

Here we also assumed that all the modes that are relevant for the CMB and structure formation have gone through all three phases of evolution of dispersion relation from the onset of inflation. This would be true if Inflation lasted for the number of e-foldings larger than the minimum number required to solve the problems of standard Big Bang cosmology.  In this case, if the largest scale mode that corresponds to our horizon today satisfies this condition, the smaller ones in the CMB have also experienced all three phases of the dispersion relation throughout their evolution. However, the reverse is not necessarily true; if the smaller modes are already in the regime of domination of the $k^6$ term, the largest mode could have started from $k^4$ or $k^2$ regions. Therefore the initial condition which should be set for these modes are not the same as \eqref{initcond2} and the amplitude of the power spectrum would be different. It would be interesting to investigate the observational consequences of such situations. This is another interesting feature that we plan to get back to in a future study. 

Here we only discussed the possibility of generating sextic modified dispersion relations with negative quartic terms from EFT of Inflation for the scalar perturbations. If similar contributions could be generated for tensor perturbations, one can expect similar enhancement for the tensor power spectrum too. Assuming that both tensor and scalar perturbations are affected similarly, that would mean that the tensor-scalar ratio, $r$, remains intact, but the scale of inflation could be lowered. Then $r\gtrsim 0.01$ would not necessarily correspond to a GUT scale inflation. Investigating whether such dispersion relations is conceivable for tensor perturbations, is another avenue to pursue. 

\section*{Acknowledgement}

We are grateful to P. Creminelli, M. M. Sheikh-Jabbari, A. Maleknejad and G. Shiu for helpful discussions.  GG and HJK research is supported by the Discovery Grant from Natural Science
and Engineering Research Council of Canada.

\bibliographystyle{JHEP}
\bibliography{bibtex}

\end{document}